\documentclass[aps,prl, twocolumn]{revtex4}
\pdfoutput=1
\usepackage{amsmath}
\usepackage{ amssymb }
\usepackage{graphicx}
\usepackage{color}

\begin{document}

\newcommand{\beq}{\begin{equation}}
\newcommand{\eeq}{\end{equation}}
\newcommand{\beqa}{\begin{eqnarray}}
\newcommand{\eeqa}{\end{eqnarray}}
\newcommand{\ben}{\begin{enumerate}}
\newcommand{\een}{\end{enumerate}}
\newcommand{\hs}{\hspace{0.5cm}}
\newcommand{\vs}{\vspace{0.5cm}}
\newcommand{\note}[1]{{\color{red} [#1]}}
\newcommand{\tim}{$\times$}
\newcommand{\bigo}{\mathcal{O}}
\newcommand{\bra}[1]{\ensuremath{\langle#1|}}
\newcommand{\ket}[1]{\ensuremath{|#1\rangle}}
\newcommand{\bracket}[2]{\ensuremath{\langle#1|#2\rangle}}
\renewcommand{\vec}[1]{\textbf{#1}}

\newcommand{\dagga}{{\phantom{\dagger}}}
\newcommand{\ud}{\,\mathrm{d}}

\newcommand{\bo}{B} 
\newcommand{\wkz}{\omega_{\vec{k},z}}
\newcommand{\qq}{q} 
\newcommand{\jp}{J'} 
\newcommand{\jpl}{J} 
\newcommand{\gk}{\gamma_{\vec{k}}}

\newcommand{\todo}[1]{{\color{red} {\bf [ToDo: #1]}}}


\title{Excitation Gap Scaling near Quantum Critical Three-Dimensional Antiferromagnets}

\author{M. Loh\"ofer}
\affiliation{Institut f\"ur Theoretische Festk\"orperphysik, JARA-FIT and JARA-HPC, RWTH Aachen University, 52056 Aachen, Germany}

\author{S. Wessel}
\affiliation{Institut f\"ur Theoretische Festk\"orperphysik, JARA-FIT and JARA-HPC, RWTH Aachen University, 52056 Aachen, Germany}

\begin{abstract}
By means of large-scale quantum Monte Carlo simulations, we examine the quantum critical scaling of the magnetic excitation gap (the triplon gap)  in a three-dimensional dimerized quantum antiferromagnet, the bicubic lattice, and  identify characteristic  multiplicative logarithmic scaling corrections atop the leading mean-field behavior.  These findings are in accord with 
field-theoretical predictions that are based on an effective description of the quantum critical system in terms of an asymptotically-free field theory,
which exhibits a logarithmic decay of the renormalized interaction strength upon approaching the quantum critical point.
Furthermore, using  bond-based singlet spectroscopy, we identify the amplitude (Higgs) mode resonance within the antiferromagnetic region. We find  a Higgs mass scaling   in accord with field-theoretical predictions that  relate it by a factor of $\sqrt{2}$ to the corresponding triplon gap in the quantum disordered regime. In contrast to the situation in lower-dimensional systems, we observe  in this three-dimensional coupled-dimer system a distinct signal from the amplitude mode  also in the dynamical spin structure factor. The width of the Higgs mode resonance  is observed to scale linearly with the Higgs mass near criticality, indicative of this critically well-defined excitation mode of the symmetry broken phase. 
\end{abstract}

\maketitle


Quantum critical three-dimensional antiferromagnets  provide  considerably valuable condensed matter realizations of  (infrared) asymptotically free quantum field theories:
based on the quantum-to-classical mapping, the 
critical field theory that describes the underlying quantum critical point is the classical  four-dimensional  O(3) $\phi^4$-theory~\cite{Chakravarty88, Sachdev10,Sachdev11,Kulik11,Oitmaa12,Scammell15}.
Due to  a logarithmic decay of the renormalized interaction strength upon approaching the critical point, this field theory exhibits logarithmic corrections to a Gaussian fixed point~\cite{ZinnJustin02, Kenna93,Kenna94, Kenna12}.
This leads to characteristic multiplicative logarithmic scaling corrections to the bare mean-field behavior in various physical quantities that are in principle accessible by  several experimental probes, such as  
in thermodynamic measurements or 
neutron and light scattering techniques~\cite{Nohadani05,Tsukamoto07,Kao13,Qin15,Scammell15}, if probed at the  relevant energy scales near the quantum critical point.

A well characterized example system of this scenario is provided by the  dimerized spin-half compound TlCuCl${}_3$:  under the application of hydrostatic pressure, this system 
features a  quantum phase transition from a gapped quantum disordered state into an antiferromagnetically ordered phase~\cite{Ruegg04}.
The magnetic excitations across the quantum critical region have been analyzed in detail recently by  inelastic neutron scattering~\cite{Ruegg05, Ruegg08,Merchant14}.  These studies identified the
evolution of the gapped magnon mode from the dimerized quantum disordered regime (frequently referred to also as the  "triplon" mode in reference to its threefold degeneracy in the isotropic Heisenberg spin-exchange case),  to the 
low-energy (transverse) Goldstone modes that accompany the spontaneous breaking of spin-rotation symmetry in the ordered phase. 
%
\begin{figure}[t]
\includegraphics[width=\columnwidth]{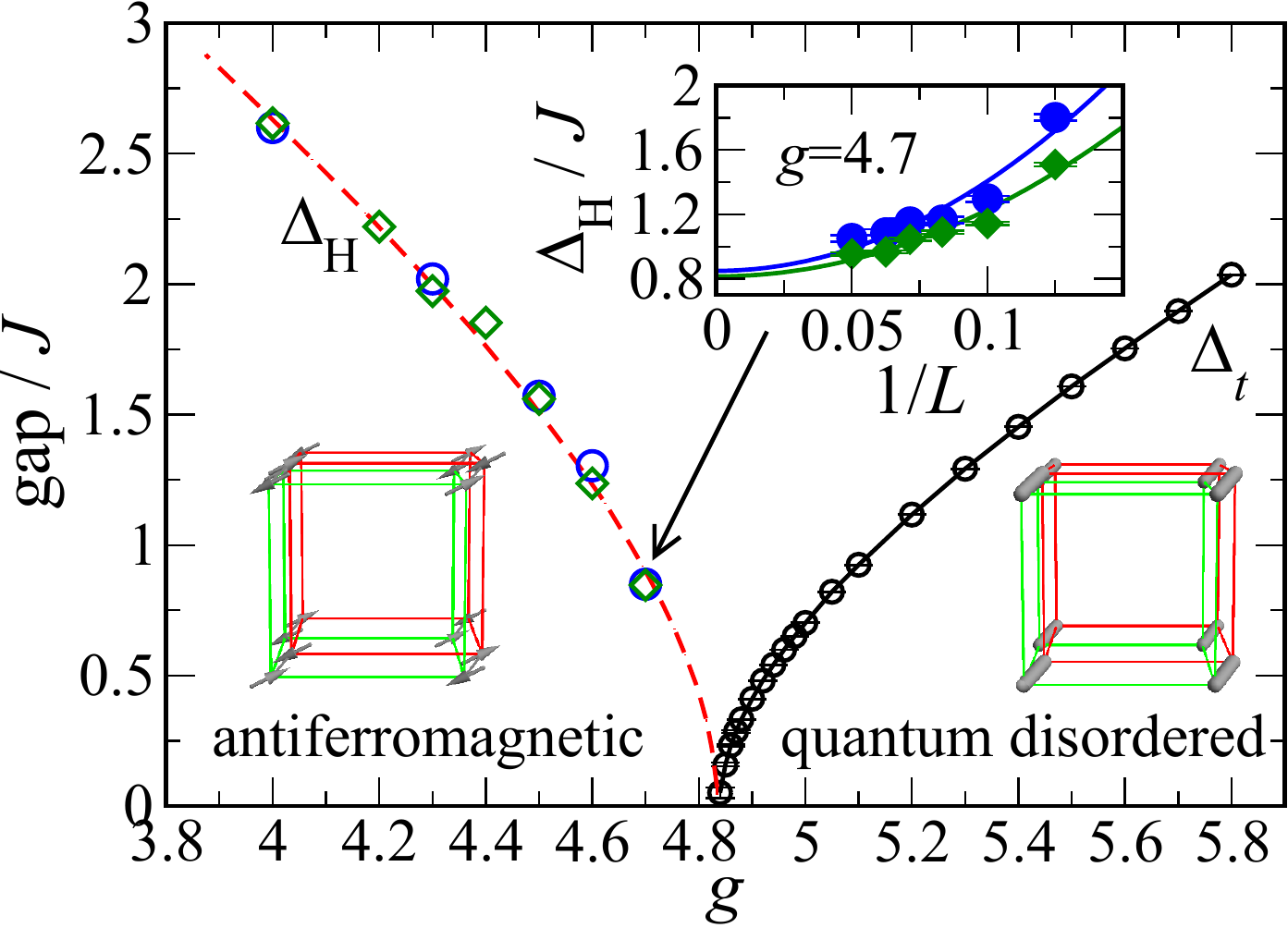}
\caption{ (Color online)
Triplon ($\Delta_t$) and Higgs ($\Delta_\mathrm{H}$)  gap in the vicinity of the quantum critical point of the bicube coupled dimer antiferromagnet. The dashed line indicates a scaling of  $\Delta_\mathrm{H}(g)=\sqrt{2}\Delta_t(g_c+(g_c-g))$. Estimates for $\Delta_\mathrm{H}(g<g_c)$ from the dynamical spin (singlet) structure factor are shown by circles (diamonds), with an estimated uncertainty of order the symbol size. The results for  
$\Delta_t(g>g_c)$ are shown by circles, and  the full line is a guide to the eye. The upper inset shows by circles (diamonds)  the finite-size values of  $\Delta_\mathrm{H}$ as estimated from the position of the second (main) peak in $S_S(\omega,0)$ ($S_B(\omega,\vec{Q})$) at $g=4.7$. Solid lines show extrapolations linear in $1/L^2$ to the thermodynamic limit.  
Both phases on the bicube lattice are  illustrated by the lower insets, where  colored (grey) bonds denote $J$ ($J'$) couplings.}

\label{fig:allgaps}
\end{figure}
%
Furthermore,  inelastic neutron scattering on this compound  also identified a gapped (longitudinal) amplitude mode of the order-parameter field~\cite{Ruegg08,Merchant14}, frequently referred to recently as a Higgs mode~\cite{Podolsky11, Pekker15}. 
This amplitude mode  softens upon approaching the quantum critical point~\cite{Affleck92,Sachdev08}.
Within a Gaussian field theory description, its excitation gap (the Higgs mass) $\Delta_\mathrm{H}(g)$  scale as 
$\Delta_\mathrm{H}(g)=\sqrt{2}\Delta(g)$ with the mass scale $\Delta(g)$ in the vicinity of the quantum critical point~\cite{Sachdev08a, Scammell15}. Here, $g$ is a  dimensionless tuning parameter (related to pressure), with the  critical point located at $g=g_c$. The mass scale $\Delta(g)$ in the antiferromagnetic region, $g<g_c$, relates via $\Delta(g)=\Delta_t(g_c+(g_c-g))$ to the triplon excitation gap $\Delta_t(g)$ in the quantum disordered regime, $g>g_c$. 
Recently, the neutron scattering data for TlCuCl${}_3$ has been re-analyzed in perspective of the asymptotic freedom scenario~\cite{Scammell15}.
It was however also  argued that the available experimental data  may not provide robust evidence for logarithmic corrections, given the size of the  error margin  and a reduced number of data points near the quantum critical point~\cite{Qin15}.
In order to probe for logarithmic scaling corrections in quantum critical spin dimer systems, 
it is  thus crucial to  compare these field-theoretical  predictions with unbiased high-precision results.

Here, we  provide such a characterisation  of the quantum critical scaling of the excitations in three-dimensional dimerized antiferromagnets by  addressing  directly the relevant dynamical quantities using quantum Monte Carlo (QMC) simulations. 
In particular, we  analyze the scaling of  the magnetic (triplon) excitation gap ($\Delta_t$) as well as the Higgs mass ($\Delta_\mathrm{H})$ near the quantum critical bicube Heisenberg model, the most basic three-dimensional coupled dimer system. 
By probing the system in close vicinity to its quantum critical point, we identify multiplicative logarithmic scaling corrections and also confirm the characteristic $\sqrt{2}$ value of the gap-ratio. 
Furthermore, we find that -- in contrast to the two-dimensional case of the Heisenberg bilayer system -- the Higgs excitation mode in the three-dimensional bicube system can be   identified not only  by the singlet-based scalar susceptibility~\cite{Matsumoto08,Podolsky11,Matsumoto14,Lohoefer15} but also as a distinct resonance mode  in the dynamical spin structure factor,
which  relates directly to  inelastic neutron scattering.  
Before presenting our findings, we first introduce the model system and the used QMC approach. 
The spin-1/2 Heisenberg model on the bicubic lattice consists of   an arrangement of spin dimers on a simple cubic  lattice:  each unit cell contains one such dimer, with a common  vector  connecting the  two spins  forming the dimer in each unit cell~\cite{Jin12,Qin15}. 
The Hamiltonian is hence given by 
\beq
\label{hh}
 H= J' \sum_i  \vec{S}_{i1} \cdot \vec{S}_{i2} + J \sum_{\langle i,j \rangle} (\vec{S}_{i1}\cdot\vec{S}_{j1}+\vec{S}_{i2}\cdot\vec{S}_{j2}),
\eeq
where spin $\vec{S}_{i\mu}$ resides on the first ($\mu=1$) and second ($\mu=2$) site  of the dimer within the $i$-th unit cell of the cubic lattice (see Fig.~\ref{fig:allgaps} for an illustration).
Furthermore, $J'$ denotes the coupling within each dimer, and $J$ the coupling between spins in different unit cells. 
In the following, we denote by $g=J'/J$ the ratio of the two coupling constants and set the lattice constant $a$ of the cubic lattice to $1$. 
For this spin dimer system, multiplicative logarithmic corrections  were 
identified in several thermodynamic quantities, such as the $g$-dependence of the ordering temperature, in the vicinity of the quantum critical point
 at $g=g_c=4.83704(6)$~\cite{Qin15} that separates the antiferromagnetic low-$g$ phase from the quantum disordered large-$g$ regime. 
The bicube model contains an inversion symmetry with respect to exchanging the spins with $\mu=1, 2$  in all unit cells. 
We account for this additional quantum number by  assigning a forth  component to an originally three-dimensional momentum space vector. Hence, $\vec{k}=(k_x,k_y,k_z,k_p)$, with $k_p=0$ or  $\pi$, denoting the symmetric and antisymmetric channel with respect to dimer inversion, respectively. 
Correspondingly, each spin is assigned a position vector
$\vec{r}_{i\mu}$, with a forth component equal to $0$ ($1$), for $\mu=1$ ($\mu=2$).

Of particular interest
to our analysis 
is the dynamical spin structure factor
$
S_S(\omega,\vec{k})= \frac{1}{N_s}\int dt\sum_{i,j,\mu,\nu} e^{i(\omega t - \vec{k}\cdot(\vec{r}_{i\mu}-\vec{r}_{j\nu}))} \langle \vec{S}_{i\mu}(t) \cdot\vec{S}_{j\nu}(0) \rangle,
$
where $N_s$ denotes the number of spins, and $k_p=0$ ($k_p=\pi$)  refers to  the 
symmetric (antisymmetric) sector.
In the presence  of long-range antiferromagnetic order, one may also  distinguish  the components  of $S_S(\omega,\vec{k})$ parallel and transverse with respect to the order parameter orientation, in which case $S_S(\omega,\vec{k})$ represents a rotational average that is probed by  the QMC simulations.
In addition to the spin correlations, we also analyse correlations among the dimer bond-based spin-exchange terms,
$
B_i=\vec{S}_{i1} \cdot \vec{S}_{i2},
$
and define a corresponding scalar response function in terms of the dynamical singlet structure factor  
$
S_B(\omega,\vec{k})= \frac{1}{N_d}\int dt\sum_{i,j} e^{i(\omega t - \vec{k}\cdot(\vec{r}_{i}-\vec{r}_{j}))} \langle B_i(t) B_{j}(0) \rangle,
$
where $N_d=N_s/2$ denotes the number of dimers.
Here, $\vec{k}$ and the $\vec{r}_i$ denote three-dimensional cubic lattice k-space and lattice position vectors (i.e., with a vanishing forth component) respectively, where  the operator $B_i$ resides at position $\vec{r}_i$ on the simple cubic lattice.  

We analyze these dynamical quantities of the bicube Heisenberg model using QMC simulations based on the stochastic series expansion method with directed loop updates~\cite{Sandvik99,Syljuasen02,Alet05}, considering finite systems with $N_s=2L^3$ lattice sites and  periodic boundary conditions.
We used systems with $L$ ranging from $8$ up to $26$ close to the quantum critical point, which corresponds to  up to $N_s=35152$ spins.
In order to access ground state properties, the inverse temperature $\beta$ has  been chosen sufficiently large. This typically required $\beta J\geq 2 L$. 
In order to  calculate the dynamical spin structure factor, we  efficiently~\cite{Michel07} measured the imaginary-time displaced spin-spin correlation functions directly in Matsubara frequency representation~\cite{SubMat}.
The numerical inversion to obtain $S_S(\omega,\vec{k})$ from the Matsubara frequency QMC data
was performed using the stochastic analytic continuation method in the formulation of Ref.~\onlinecite{Beach04}. 
For the dynamical singlet structure factor $S_B(\omega,\vec{k})$, we
measured  the corresponding bond-bond correlation functions directly in imaginary-time, binned over finite-width imaginary-time windows~\cite{Michel07}. Using an appropriate kernel for the analytic continuation, we can directly relate $S_B(\omega,\vec{k})$ to these imaginary-time binned data~\cite{Lohoefer15}.


The  dynamical spin structure factor 
 $S_S(\vec{k},\omega)$ is dominated by the single-magnon dispersion, which for $g<g_c$ softens at the antiferromagnetic Bragg peak position  $\vec{Q}=(\pi,\pi,\pi,\pi)$, while a finite triplon gap $\Delta_t$ exists at $\vec{k}=\vec{Q}$  in the structure factor data for $g>g_c$~\cite{SubMat}.
For a  quantitative analysis of the $g$-dependence of $\Delta_t$ in the thermodynamic limit, we performed a systematic finite-size scaling analysis. For this purpose, we  obtained the values of  $\Delta_t$ for various $g>g_c$ and different system sizes by extracting from the imaginary-time spin-spin correlation function $S_S(\tau,\vec{k})$ at $\vec{k}=\vec{Q}$ the low-temperature asymptotic form $S_S(\tau,\vec{Q})\propto \exp{(-\tau \Delta_t)}$. Here, $S_S(\tau,\vec{k})$ is obtained from the Matsubara frequency data  by a discrete Fourier back-transformation~\cite{Pollet12}.
%
Based on the finite-size dependence $\xi_\tau(L)=\xi_\tau-b \exp{(c L)}$ of the correlation-length in the imaginary-time 
direction, where $\xi_\tau=1/\Delta_t$~\cite{Matsumoto01,SubMat}, and $b$ and $c$ are fit parameters, 
we obtain the thermodynamic limit values of  $\Delta_t$ (cf. Fig.~\ref{fig:allgaps}).

\begin{figure}[t]
\includegraphics[width=\columnwidth]{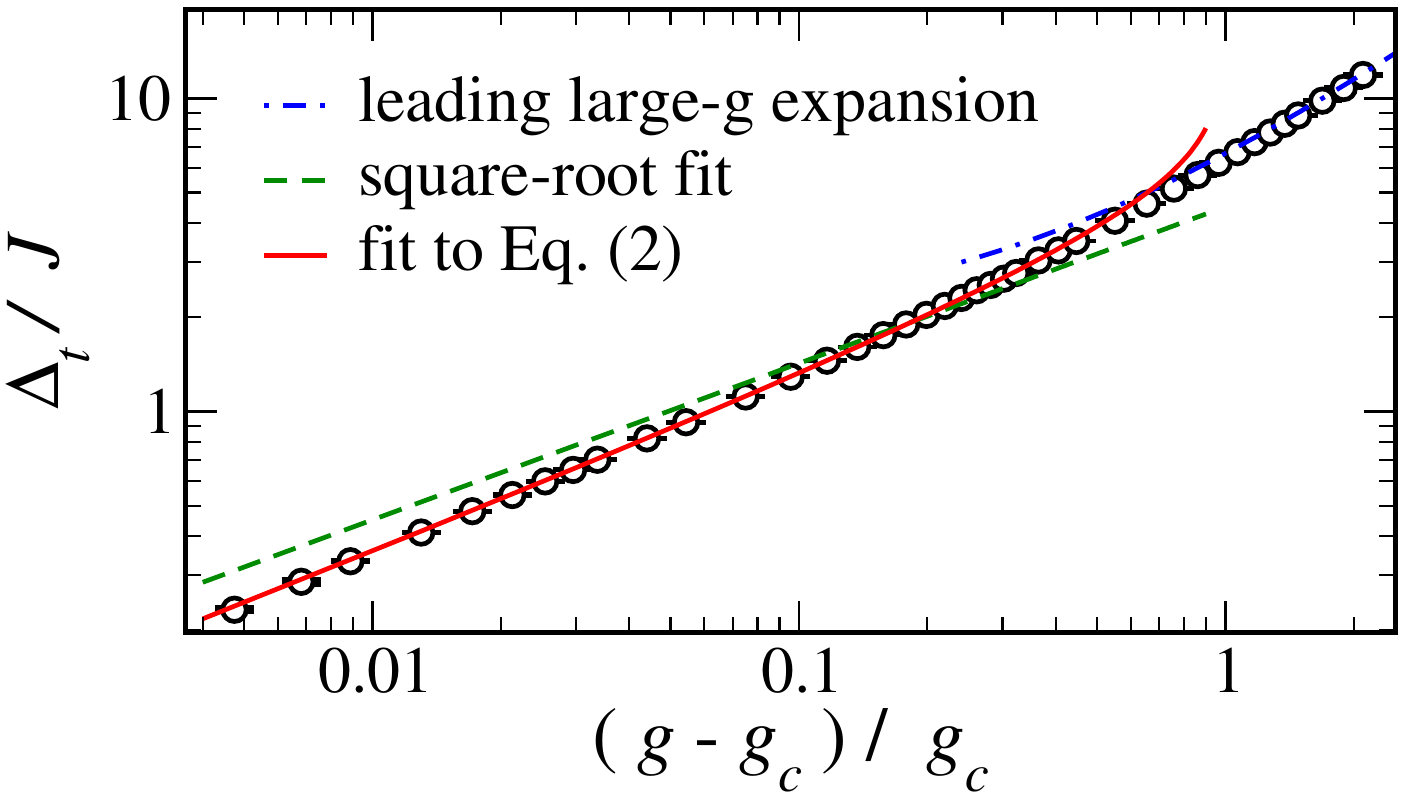}
\caption{(Color online) 
Quantum critical scaling of the triplon excitation gap $\Delta_t$ in the vicinity of the quantum critical point. The QMC estimates, obtained after a finite-size extrapolation, are shown by black circles. The solid line is a fit to Eq.~(\ref{eq:tglog}) of the data with $(g-g_c)/g_c<0.3$,  the dashed line indicates  Gaussian theory square-root scaling,
and the dashed-dotted line  the  leading large-$g$ perturbative expansion result.}
\label{fig:tglog}
\end{figure}

In order to  closer examine the quantum critical scaling, $\Delta_t$  is shown as a function of the relative distance from the quantum critical point, $(g-g_c)/g_c$, in Fig.~\ref{fig:tglog} in a log-log plot.  
Also included in Fig.~\ref{fig:tglog} is a fit of the data to a square-root scaling  proportional to $(g-g_c)^{1/2}$, corresponding to  Gaussian mean-field behavior. 
It is clear from Fig.~\ref{fig:tglog}, that this scaling form does not account for the gap data in the critical region. In fact,  the  
logarithmic decay of the renormalized interaction strength upon approaching the quantum critical point leads to a logarithmic correction to the mean-field scaling behavior,
\begin{equation}\label{eq:tglog}
\Delta_t(g)\propto \left(\frac{g-g_c}{g_c}\right)^{1/2} \left|\ln\frac{g-g_c}{g_c}\right|^{-5/22},
\end{equation}
in the vicinity of the quantum critical point~\cite{Kenna12, Kenna13,Scammell15,Coester16}. This follows from the quantum-to-classical mapping with a dynamical critical exponent $z=1$,  and relating $\Delta_t=\xi_\tau^{-1}$ to the correlation length in the imaginary-time direction. The solid line in Fig.~\ref{fig:tglog} shows that the numerical data  is well in accord with  this analytic prediction up to  values $(g-g_c)/g_c\lesssim 0.3$, thus setting the critical region. Fitting the numerical data to the scaling law in Eq.~(\ref{eq:tglog}) with the  exponent $5/22$ replaced by a free 
fit-parameter $\bar{\nu}$, we obtain an independent estimate of $\bar{\nu}=5.2(1)/22$,
when all data with $(g-g_c)/g_c< 0.25$ is included,  and with a reduced $\chi^2$ value (per degree of freedom)  of $\chi^2/d.o.f.=1.2$. 
Further analysis shows that the triplon gap data does not fit well to such a scaling form anymore (with 
values of $\chi^2/d.o.f.$ rapidly exceeding values of order 10), 
if further data points beyond $(g-g_c)/g_c>0.35$ are included into the fit range. 
 Also shown in  Fig.~\ref{fig:tglog}  
 is the  large-$g$ perturbative expansion result $\Delta_t=J'-3J + O(J^2)$ for the triplon gap, which traces the data well for $(g-g_c)/g_c>1$.  
 We note that in the crossover region 
 no indication for a distinct pure Gaussian mean-field behavior can be seen.
 

\begin{figure}[t]
\includegraphics[width=\columnwidth]{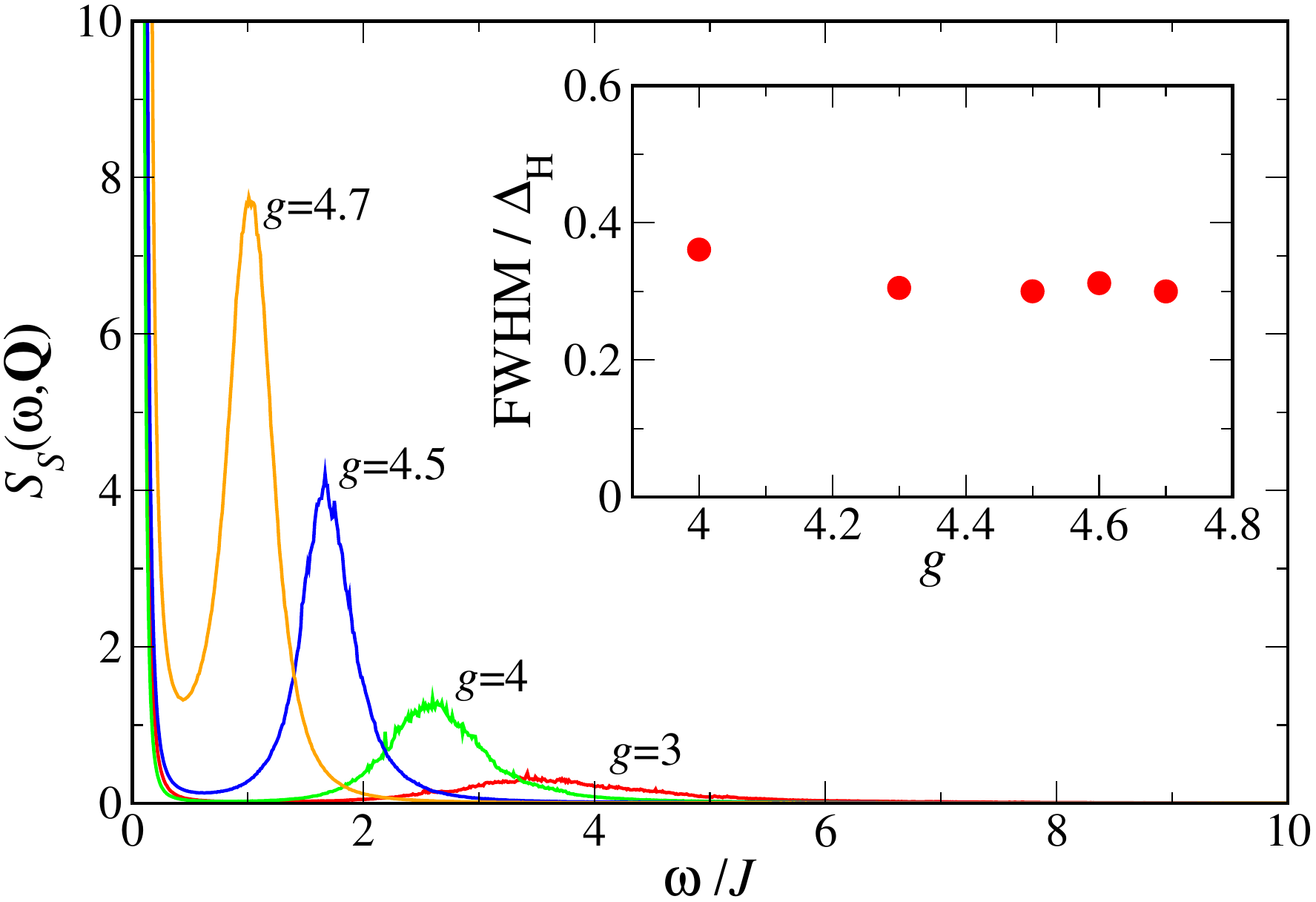}
\caption{(Color online) 
Dynamical spin structure factor $S_S(\omega,\vec{Q})$ at the ordering wave vector for different values of the coupling ratio $g$, as obtained from QMC simulations with $L=20, \beta=40/J$. 
The inset shows of the ratio between the width (FWHM) and the position $\Delta_\mathrm{H}$ of the second  peak, where the estimated uncertainty is of order the symbol size. }
\label{fig:L20PiPiPiPi}
\end{figure}

We next address the amplitude (Higgs) mode in the bicube model. 
Fig.~\ref{fig:L20PiPiPiPi} shows 
 the spin spectral function $S_S(\omega,\vec{Q})$ for various values of $g<g_c$. 
%
%
While the signal is dominated by the Bragg peak at $\omega=0$, we  also identify a second, distinct spectral  feature for all shown values of $g$. 
This  broader peak  furthermore softens upon approaching the quantum critical point and also sharpens closer to $g_c$. 
%
%
By tracing the peak position as a function of $g$, we obtain the excitation energies indicated by open circles in Fig.~\ref{fig:allgaps}, after performing again an extrapolation to the thermodynamic limit (an example of which is shown in the inset of Fig.~\ref{fig:allgaps} for $g=4.7$). 
Also included in  Fig.~\ref{fig:allgaps} is the field-theory prediction~\cite{Sachdev08a, Scammell15} for the Higgs mass scaling $\Delta_\mathrm{H}(g)=\sqrt{2}\Delta(g)$,
with $\Delta(g)=\Delta_t(g_c+(g_c-g))$. 
The extrapolated  peak positions closely follow this scaling prediction, indicating that this second feature in $S_S(\omega,\vec{Q})$ indeed signals the amplitude mode of the bicube system. One may compare this to the case of the two-dimensional bilayer model, where the amplitude mode's contribution to the spin spectral function is  masked by a broader  tail atop the Goldstone mode~\cite{Zwerger04,Podolsky11,Lohoefer15}. Here, in the three-dimensional bicube system, we can clearly identify the amplitude  mode in the dynamical spin structure factor. This observation is  well in accord with the  identification of a broad amplitude mode and its softening near criticality in the neutron scattering data~\cite{Ruegg08,Merchant14} on TlCuCl${}_3$. 

\begin{figure}[t]
\includegraphics[width=\columnwidth]{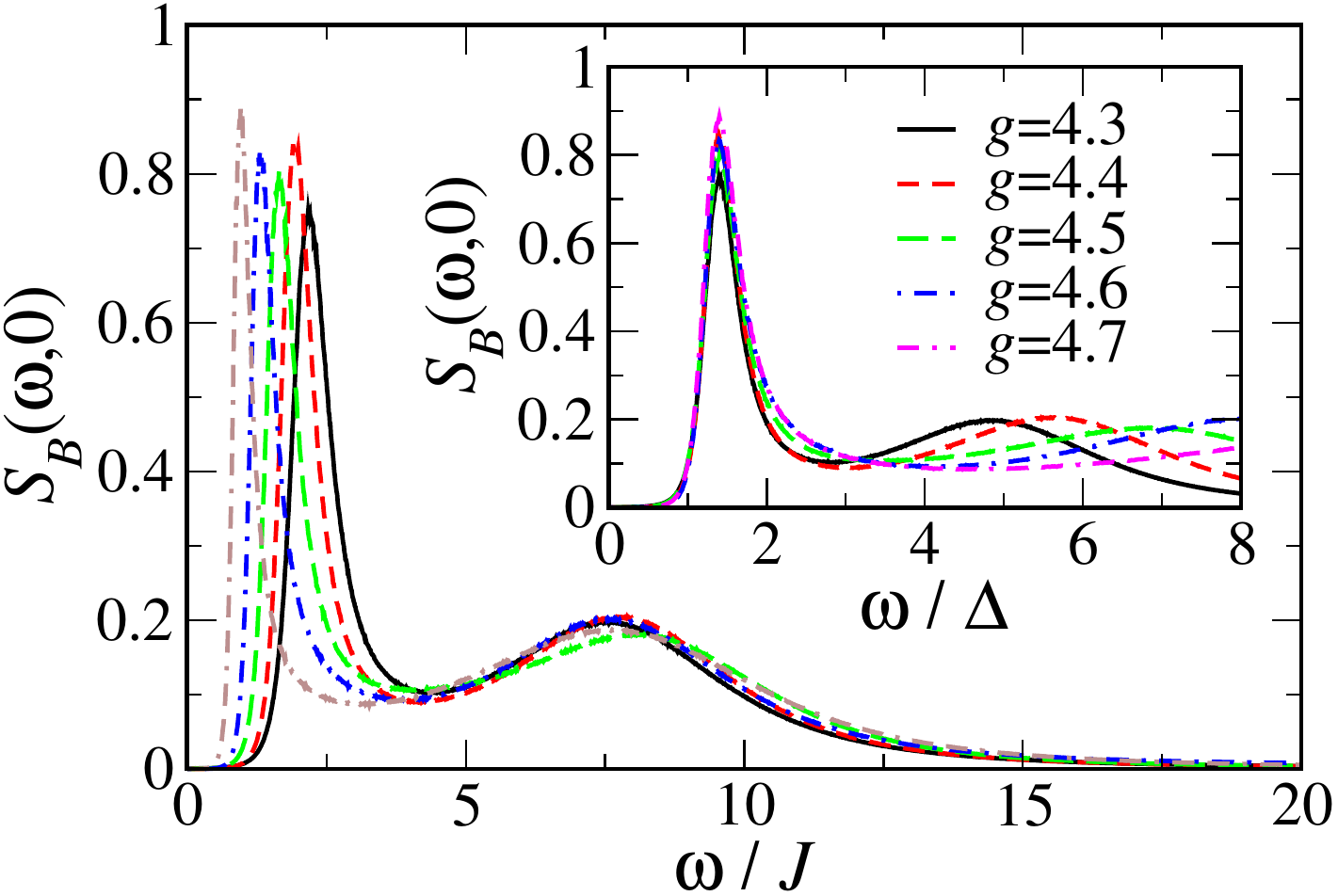}
\caption{(Color online) 
Dynamical singlet  structure factor $S_B(\omega,0)$  for different values of the coupling ratio $g$, as obtained from QMC simulations with $L=16, \beta=32/J$. The inset shows  a collapse of the Higgs peak  after rescaling of the $\omega$-axis by the energy scale $\Delta=\Delta_\mathrm{H}/\sqrt{2}$. }
\label{fig:bondspectra}
\end{figure}

We can access the amplitude mode also from the dynamical singlet structure factor $S_B(\omega,0)$.
Due to its scalar character, this quantity contains the amplitude-mode signal without being masked by the low-energy Goldstone 
 modes~\cite{Matsumoto08,Podolsky11,Matsumoto14,Lohoefer15}. This fact was employed in Ref.~\onlinecite{Lohoefer15} in order to access the amplitude mode for the two-dimensional bilayer system. Here, we perform an  analysis of $S_B(\omega,0)$ for the bicube model:
Figure~\ref{fig:bondspectra} shows  $S_B(\omega,0)$  for various values of $g$. The amplitude mode dominates $S_B(\omega,0)$  as a pronounced low-energy peak that softens upon approaching the quantum critical point. It is followed 
by a second, broader peak at more elevated energies of $\omega/J\approx 7.5$, which exhibits no such clear $g$-dependence, and thus does not relate to the critical low-energy spectrum. 
From  an  extrapolation of the main peak's position to the thermodynamic limit, we extract the Higgs mass scaling in the vicinity of the quantum critical point 
(this extrapolation is also shown explicitly for $g=4.7$ in the inset of Fig.~\ref{fig:allgaps}).
The resulting $g$-dependence of the Higgs mass $\Delta_g$ is  shown in the main panel of  Fig.~\ref{fig:allgaps}. 
We find that the extrapolated values of the main peak positions in $S_B(\omega,0)$  agree remarkable well with the extrapolated values for $\Delta_\mathrm{H}$ obtained from $S_S(\omega,\vec{Q})$. 
%
Hence, 
both the low-energy peak in  $S_B(\omega,0)$ as well as the finite energy peak in  $S_S(\omega,\vec{Q})$ relate to the amplitude mode in the bicube system, with an excitation energy that (i) softens upon approaching the quantum critical point and (ii)  exhibits a $g$-dependence in accord with the field-theory prediction, 
$\Delta_\mathrm{H}=\sqrt{2}\Delta$~\cite{Sachdev08a, Scammell15}.

Even though the form of the Higgs peak is affected  by the analytical continuation procedure, we 
find that 
our numerical results for the overall shape of the Higgs peak compare well  to the universal low-energy scaling form 
$S_B(\omega,0)=\Delta^{d+z-2/\nu}\Phi(\omega/\Delta)$
of the scalar response function from Ref.~\onlinecite{Podolsky12}.
Here, $d$ denotes the spatial dimension, $z$ the dynamical critical exponent, and $\nu$ the correlation length exponent of the quantum critical point. Finally, $\Phi$ is a corresponding scaling function.
For the bicube system, where $d=3$, $z=1$, and $\nu=1/2$, we simply obtain $S_B(\omega,0)=\Phi(\omega/\Delta)$. This is in good accord to the overall collapse of the Higgs peak signals shown in the inset of Fig.~\ref{fig:bondspectra}, in particular in the low-$\omega$ section. It furthermore shows  that the Higgs peak sharpens upon approaching the quantum critical point with an only weakly $g$-dependent ratio between its width and the Higgs mass.  
Within the accuracy for the peak width that is 
available by  the analytic continuation scheme, we estimate this ratio $R$ from the inset of Fig.~\ref{fig:bondspectra} in terms of its full width at half maximum (FWHM), to $R=0.4(1)$. A similarly roughly constant  ratio between the FWHM and the peak position ($\Delta_\mathrm{H}$) is also obtained for the Higgs peak signal in the dynamical spin structure factor (cf. the inset of Fig.~\ref{fig:L20PiPiPiPi}).  Both findings establish the amplitude mode as a relatively broad, critically well-defined excitation on the bicube lattice, as observed also in experiments~\cite{Ruegg08,Merchant14} on TlCuCl${}_3$ and within field-theory calculations~\cite{Kulik11, Affleck92}.

In summary, we provided robust evidence for the presence of multiplicative logarithmic corrections to the leading Gaussian mean-field scaling of the triplon excitation gap softening near the  quantum critical point. We 
observed a distinct signal of the amplitude mode in both the  spin and the dimer bond-based singlet dynamical structure factors. This Higgs mode furthermore softens and sharpens upon approaching the quantum critical point,
with an excitation gap  that scales consistently with a $\sqrt{2}$-ratio to the corresponding triplon excitation energy. The rather broad width of the Higgs peak in both quantities scales essentially proportional with the Higgs mass. These findings close the gap between the field-theoretical description of these fundamental quantum phase transitions and their experimental investigation in three-dimensional  dimerized antiferromagnets.

\subsection*{Acknowledgments}
We thank F. Mila
for discussions and Z.Y. Meng for mentioning related findings~\cite{Quin16}.
We acknowledge support by the Deutsche Forschungsgemeinschaft (DFG) under grant FOR 1807.
We also  thank the IT Center at RWTH Aachen University and the JSC J\"ulich for access to computing time through JARA-HPC.
%


%
%

\clearpage
\section{Supplemental Material}
\setcounter{figure}{0}  
\renewcommand{\thefigure}{S\arabic{figure}}

\begin{figure*}[t]
\includegraphics[width=2\columnwidth]{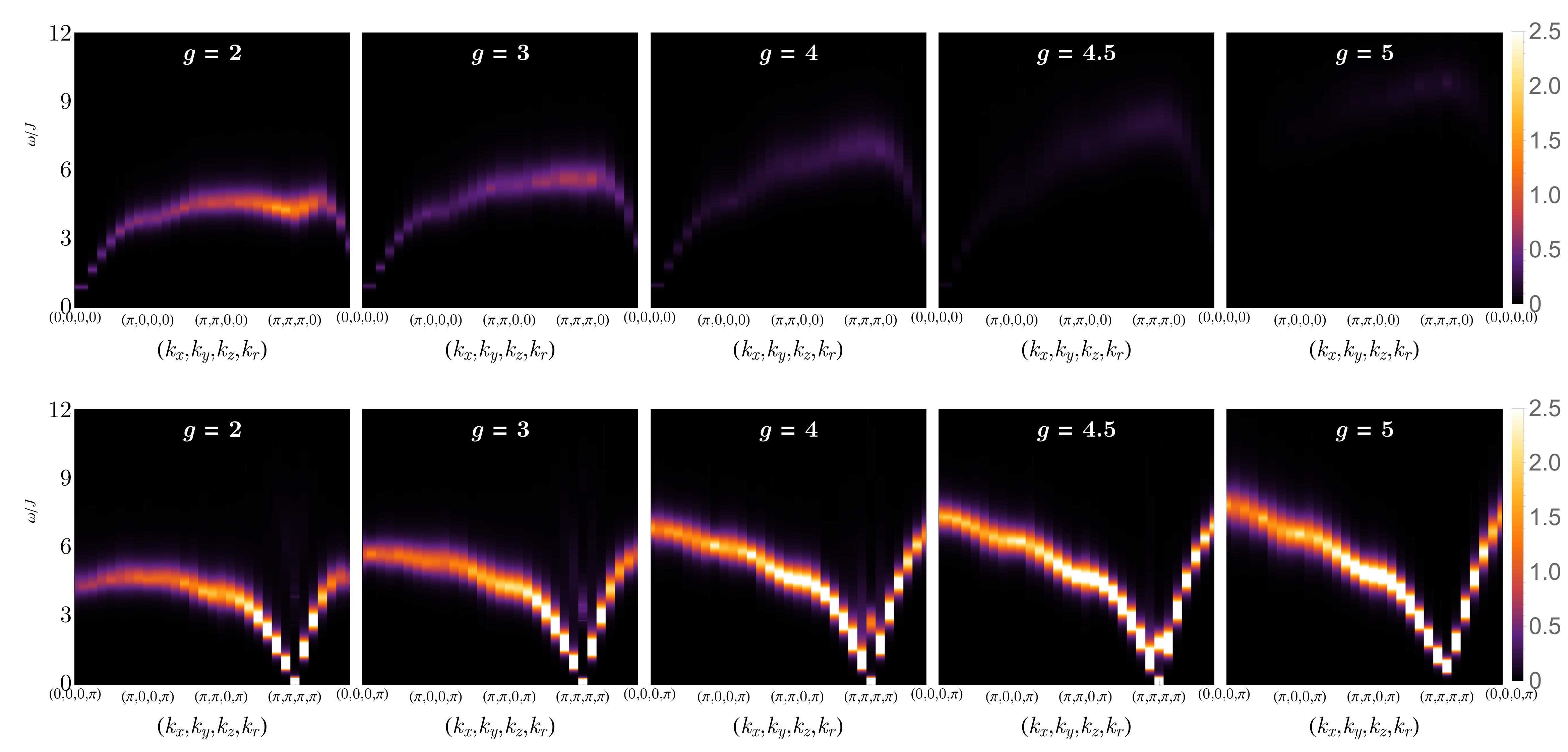}
\caption{(Color online) 
Dynamical spin structure factor $S_S(\omega,\vec{k})$ in the symmetric (upper panel) and antisymmetric (lower panel) channel for the spin-half Heisenberg model on the bicubic lattice at different ratios $g=J'/J$ of the intradimer ($J'$) to interdimer  ($J$) coupling strength along the indicated path in the three-dimensional Brillouin zone. The spectral functions shown in this figure were obtained from an analytic continuation based on Matsubara frequency data for a $L=20$ system at $\beta/J=40$.}
\label{fig:SS}
\end{figure*}

\begin{figure}[t]
\includegraphics[width=\columnwidth]{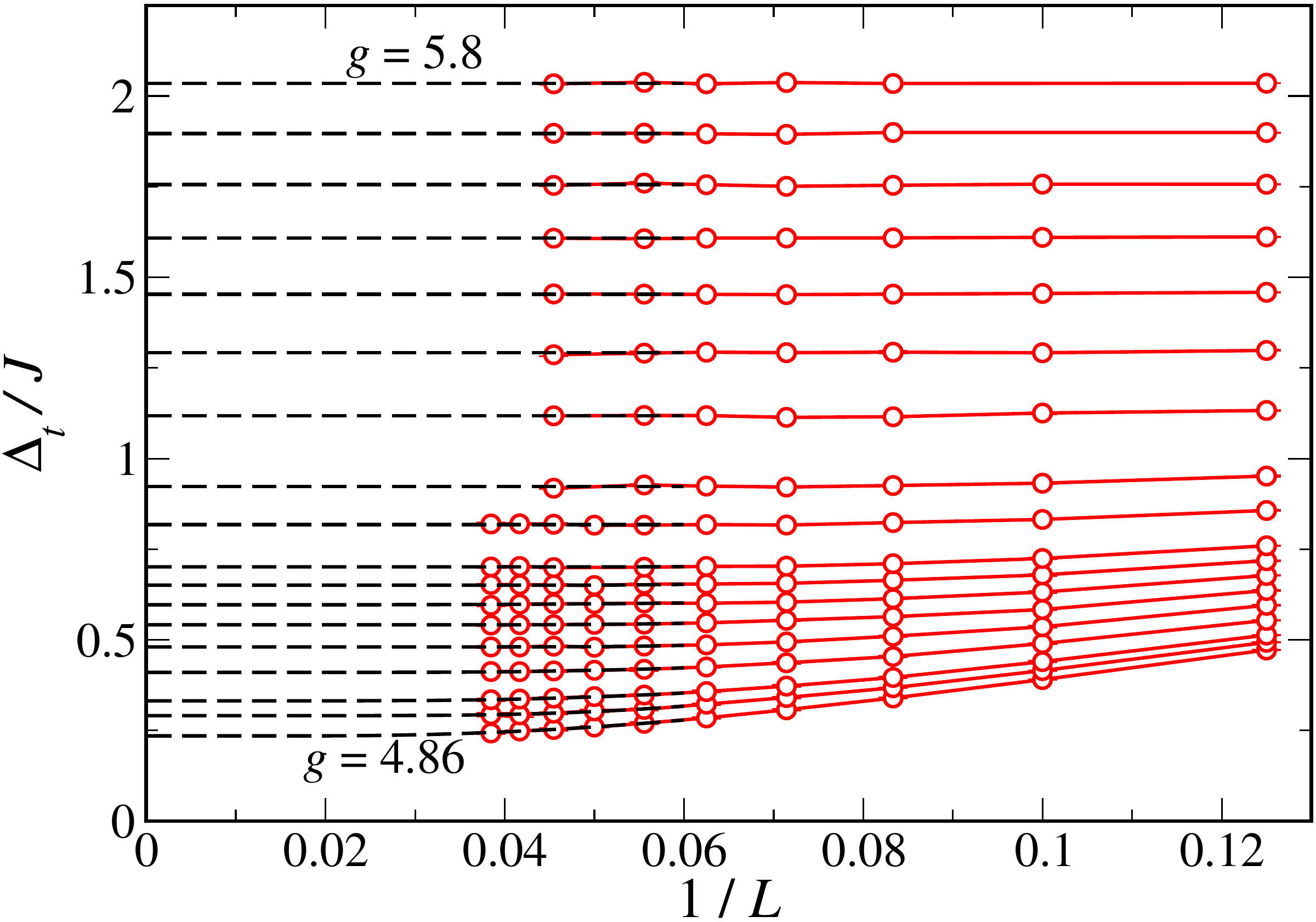}
\caption{(Color online) 
Finite-size scaling of the triplon excitation gap $\Delta_t$ close to the quantum critical point for different values of $g=J'/J$ inside the quantum disordered regime. Results are shown for $g=4.86$,   
$4.87$, 
$4.88$, 
$4.9$, 
$4.92$, 
$4.94$, 
$4.96$, 
$4.98$, 
$5$, 
$5.05$, 
$5.1$, 
$5.2$, 
$5.3$, 
$5.4$,
$5.5$, 
$5.6$, 
$5.7$, 
$5.8$
(bottom to top). 
Dashed lines indicate the finite size extrapolation lines (cf. the main text). }
\label{fig:tgfss}
\end{figure}

The symmetric and antisymmetric structure factors as introduced  in the main text  are
related to the neutron scattering intensity  for 
a general three-dimensional bicube system with a dimer vector $\vec{d}$, which connects the two spins in each unit cell, and a general value of  the lattice constant $a$ as follows:
The scattering intensity at a scattering wave vector $\vec{q}=(q_x,q_y,q_z)$
is  proportional to 
\beqa
S^{\mathrm{tot}}_S(\omega,\vec{q})&=&\cos^2( \vec{d} \cdot \vec{q}/2 )\: S_S(\omega,a \: q_x,a \: q_y,a \: q_z, 0)\nonumber \\
&+&\sin^2(\vec{d} \cdot \vec{q}/2)\: S_S(\omega,a\:  q_x,a \: q_y,a \: q_z,\pi).\nonumber
\eeqa
We thus focus here on the two symmetry-projected sectors, since the  scattering intensity may be constructed  from these by the above relation. 

As mentioned in the main text, we measured in the QMC simulations the imaginary-time spin structure factor directly in Matsubara frequency space, 
which are related to $S_S(\omega,\vec{k})$ via
\beq\label{kernel}
S_S(i\omega_n,\vec{k})=\int_0^\infty {d\omega}\: \frac{\omega}{\pi} \frac{(1-e^{-\beta\omega})}{\omega^2+\omega_n^2}\:  S_S(\omega,\vec{k}).\nonumber
\eeq
Here,  $\omega_n= 2\pi n /\beta$ for $n=0,1,2,...$  are the bosonic Matsubara frequencies, and values of $n$ up to 250 are typically required to access the leading $1/\omega_n^{2}$ asymptotic behaviour of  $S_S(i\omega_n,\vec{k})$. As also mentioned in the main text, we then obtain from this data of $S_S(i\omega_n,\vec{k})$ the dynamical spin structure factor 
$S_S(\omega,\vec{k})$ from an analytical continuation.

Our  results for the dynamical spin structure factor, obtained from  QMC simulations for different values of $g$, are summarized in Fig.~\ref{fig:SS}.  
The spectral weight in the
antisymmetic channel (shown in the lower panel)  is dominated by the single-magnon dispersion, which for $g<g_c$ softens at the antiferromagnetic Bragg peak position  $\vec{Q}=(\pi,\pi,\pi,\pi)$, while a finite triplon gap $\Delta_t$ is visible at $\vec{k}=\vec{Q}$  in the structure factor data for $g>g_c$. In the main text,  we  analyze in more detail the scaling of this triplon gap in the vicinity of the quantum critical point. The upper panel of  Fig.~\ref{fig:SS} shows the symmetric channel of the dynamical spin structure factor, which exhibits a substantial broadening of the spectral weight upon increasing $g$ along with an overall loss of the spectral weight in the quantum disordered region. These qualitative findings are similar to the behavior observed in the two-dimensional bilayer system, and  Ref.~\onlinecite{Lohoefer15} assessed  in detail how well  spin-wave theory and a $1/d$-expansion in terms of dimer bond-operators~\cite{Joshi15a, Joshi15b} (with $d$ denoting the system's dimensionality) account for these spectral properties. Due to  the higher dimensionality of the bicube system, we anticipate here an even better agreement between these analytical approaches and the numerical spectral functions. However, both approximate approaches cannot account for the logarithmic corrections that are observed  atop the leading Gaussian critical behavior in the scaling properties of the excitation gaps, which we analyzed in the main text. 
In Fig.~\ref{fig:tgfss}, we show explicitly the finite size scaling of the  triplon excitation gap $\Delta_t$ close to the quantum critical point for different values of $g=J'/J$ inside the quantum disordered regime.

\end{document}